\documentclass[conference,letterpaper]{IEEEtran}
\usepackage{amsmath,amsfonts}
\usepackage{algorithmic}
\usepackage{algorithm}
\usepackage{array}
\usepackage[caption=false,font=normalsize,labelfont=sf,textfont=sf]{subfig}
\usepackage{textcomp}
\usepackage{stfloats}
\usepackage{verbatim}
\usepackage{graphicx}
\usepackage{cite}
\usepackage{makecell}
\usepackage{booktabs}
\begin{document}

\title{Machine Learning Assisted Inverse Design of Pixelated mmWave Patch Antennas}

\author{\IEEEauthorblockN{
Nadeem Rather \IEEEauthorrefmark{2}, 
Holger Claussen \IEEEauthorrefmark{2}\IEEEauthorrefmark{3}\IEEEauthorrefmark{4}
Lester Ho \IEEEauthorrefmark{2},
 }

\IEEEauthorblockA{\IEEEauthorrefmark{2} Tyndall National Institute, Dublin, Ireland;}
\IEEEauthorblockA{\IEEEauthorrefmark{3} University College Cork, Ireland;} \IEEEauthorblockA{\IEEEauthorrefmark{4} Trinity College Dublin, Ireland.}
\{nadeem.rather, holger.claussen, lester.ho\}@tyndall.ie\vspace{-2em} 
}
\maketitle

\begin{abstract}
In this paper, a machine learning-assisted framework for
the inverse design of pixelated millimetre-wave patch antennas
targeting the 22--30\,GHz band is presented.
The antenna surface is represented as a $19{\times}23$ binary
pixel grid on a Rogers RT/duroid\,5880 substrate, where each
pixel is either metal or empty, with a continuous electrical
path from the feed enforced by design.
An initial dataset of approximately 6{,}000 full-wave CST
simulations was collected from structured-random pixel
patterns, of which only around 40\% achieved a resonance with
$|S_{11}| \leq -10$\,dB anywhere in the band, resulting in an imbalanced dataset.
To improve simulation efficiency, an XGBoost binary classifier
was trained on this data to distinguish resonant from
non-resonant patterns before simulation.
Using the classifier as a pre-simulation filter, an additional
4{,}000 patterns were selected and simulated, raising the
overall proportion of resonant designs in the combined
10{,}000-sample dataset from approximately 40\% to 52\%.
A hybrid CNN--BiLSTM forward surrogate was then trained on
this augmented dataset to predict the full complex $S_{11}$
response across 801 frequency points, using a physics-guided
composite loss that explicitly emphasises resonance dip
accuracy.
Finally, an inverse design model was developed that optimises
in a compact 64-dimensional latent space using gradient
descent to generate pixel patterns matching a desired $S_{11}$
specification.
The results show good agreement between the surrogate-predicted
and CST-simulated $|S_{11}|$ responses for the generated
designs and demonstrate the feasibility of automatically designing and reconfiguring antenna structures.
\end{abstract}


\vspace{-3mm}

\section{Introduction}
\vspace{-1mm}

\IEEEPARstart{T}he global deployment of fifth-generation (5G)
wireless networks has increased the demand for compact planar
antennas that operate reliably across the millimetre-wave
(mmWave) spectrum~\cite{shariff2022array,rappaport2013millimeter,zhang20165g,mohammed2019review}.
Among the available antenna technologies, the microstrip patch
antenna remains widely used because of its low profile, ease of
fabrication, and compatibility with integrated circuits.
A conventional microstrip antenna consists of four main
elements: the radiating patch, substrate, ground plane, and
feed mechanism.
The patch is commonly implemented using standard geometries
such as rectangular, circular, triangular, or elliptical
shapes, which simplify analysis and fabrication while providing
acceptable radiation performance. However, once the substrate and feed structure are fixed, these conventional shapes offer limited geometric flexibility and
restrict the exploration of alternative radiating topologies. 
\begin{figure}[httb]
  \centering
  \includegraphics[width=\columnwidth]{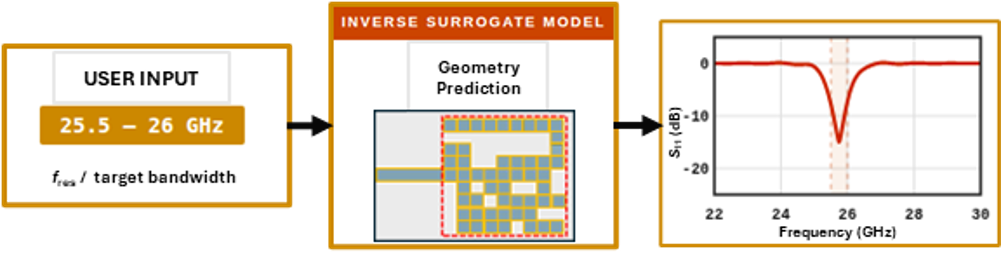}
  \caption{Overview of an AI-assisted inverse patch antenna design process}
  \label{P1.png}
\end{figure}
To address this limitation, topology optimisation methods have
been introduced in which the radiating surface is discretised
into a grid of conductive and non-conductive cells.

These \emph{pixelated} antenna structures convert the design
problem into a high-dimensional combinatorial optimisation
problem with binary design
variables~\cite{qiu2019deep, jacobs2021accurate,ghadimi2020systematic}.

\par Pixel-based antenna synthesis enables the exploration of a
substantially larger design space and can reveal
non-conventional geometries with improved impedance matching,
bandwidth, or radiation characteristics.
However, the number of possible configurations increases
exponentially with the number of pixels.
For example, a grid containing 300 pixels corresponds to a
search space of $2^{300}$ possible layouts, making direct
exhaustive evaluation impractical.
Accurate evaluation of possible antenna layouts generally
requires full-wave electromagnetic (EM) simulation using tools
such as CST Microwave Studio or Ansys HFSS \cite{CST}.
Although these simulators provide reliable characterisation of
antenna performance, each simulation may require several
minutes, making large-scale design exploration computationally
demanding. 
To reduce this cost, surrogate modelling and machine learning
(ML) techniques have been adopted to approximate the
relationship between antenna geometry and EM
response~\cite{li2022xgboost,mahouti2019design,dwivedi2025taguchi,torun2021deeplearning,chen2025inverse,koziel2023lowcost,sharma2020ml}.

Once trained, these models can provide rapid predictions and
thereby reduce the number of expensive EM simulations required
during design optimisation.
The role of ML-assisted methods in antenna design has also
been emphasised in recent surveys on 5G microstrip antenna
development \cite{comprehensive2024}.

For example, in~\cite{mahouti2019design}, a multilayer perceptron was
trained using EM simulation data to approximate the $S_{11}$
response of a reconfigurable microstrip antenna and support
faster optimisation.
\par
Similarly, Taguchi-assisted neural network models have been
used to predict the reflection coefficient of mmWave antennas
while reducing the number of repeated full-wave simulations
required during parameter tuning~\cite{dwivedi2025taguchi}.
More recently, deep learning approaches have shown improved
performance for high-dimensional antenna representations.
In particular, convolutional neural networks (CNNs) are well
suited to pixel-based antenna layouts because they can extract
spatial features directly from binary patterns.
In \cite{torun2021deeplearning} an accurate forward and inverse
design for pixelated mmWave antennas is developed using deep convolutional
models trained on large
datasets. Furthermore, in \cite{chen2025inverse} a
combined CNN-based surrogate with binary particle swarm
optimisation for pixel antenna design is reported with accurate
prediction of reflection coefficients for optimised
layouts.

Despite these advances, efficient generation of training data
remains a major challenge for surrogate-based design of
pixelated antennas.
Randomly generated pixel layouts rarely exhibit acceptable
antenna performance, which leads to highly imbalanced datasets
in which only a small fraction of samples produce meaningful
resonance behaviour.
In~\cite{chen2025inverse}, only 12.6\% of over
150{,}000 randomly generated 10$\times$10 pixel layouts produced a usable
resonance, highlighting the scale of this inefficiency.
As a result, simulation resources are often spent on designs
with limited value for surrogate training.

In this paper, a novel end-to-end machine-learning assisted
design pipeline for pixelated mmWave patch antennas in the
22--30\,GHz band is presented.
The proposed approach integrates ML-assisted dataset
augmentation, a hybrid CNN--BiLSTM forward surrogate for
complex $S_{11}$ prediction, and latent-space inverse design
using a single pipeline.

\section{Antenna Structure and Dataset}
\label{sec:structure}

\begin{figure}[!t]
  \centering
  \includegraphics[width=\columnwidth]{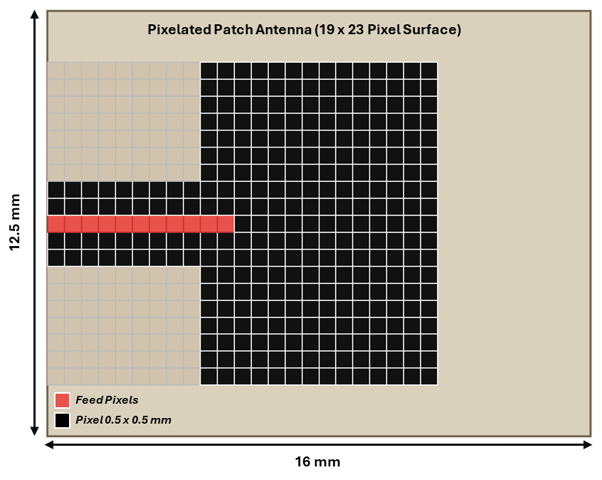}
  \caption{Full pixelated patch antenna surface on Rogers
           RT/duroid\,5880 ($16{\times}12.5$\,mm,
           $\varepsilon_r{=}2.2$, $h{=}0.508$\,mm).
           All 300 optimisable metal pixels are shown in black.
           The fixed feed line (cols.\,1--11, row\,10) is shown
           in red. Grey regions are structurally excluded.}
  \label{fig:geometry}
\end{figure}

The pixelated antenna is developed on a grounded Rogers
RT/duroid\,5880 substrate with relative permittivity
$\varepsilon_r{=}2.2$, loss tangent $\tan\delta{=}0.0009$,
and thickness $h{=}0.508$\,mm.
The substrate dimensions are $16{\times}12.5$\,mm$^2$ with a
copper thickness of $t{=}0.035$\,mm.
The top surface consists of $0.5{\times}0.5$\,mm$^2$ square
copper pixels with zero gap, forming a $19{\times}23$ grid.
Structural margins of 1.5\,mm at the upper and lower edges and
4.5\,mm at the right edge confine the active pixel area to
$19{\times}23{=}437$ sites.
A fixed feed line occupying columns\,1--11 of the centre row
(row\,10 of\,19) is kept fixed. Furthermore, to retain a patch shape some of the pixels within the pixelated surface area are excluded leaving 300 binary-valued
pixels available for optimisation.
The resulting design space spans approximately $2^{300}$
possible geometries. Fig.~\ref{fig:geometry} illustrates the full pixelated surface
with all pixel states visible.

A key challenge in building a useful training dataset for this
problem is that purely random binary pixel patterns tend to
produce geometries with disconnected metallic islands that have
no electrical path to the feed and therefore cannot radiate.
To address this, a structured-random generation strategy was
adopted, where patterns are drawn from seven predefined shape
families: \textit{rect\_block}, \textit{rect\_ring},
\textit{split\_ring}, \textit{symmetric\_mirror},
\textit{stub\_patch}, \textit{scatter\_connected}, and
\textit{inset\_patch}.
Each family defines a broad antenna topology class, with the
geometry within each class fully randomised across anchor
position, patch dimensions, growth direction, and pixel
density, providing wide coverage of each topological region of
the design space. Two constraints are enforced on every generated pattern: the
total number of active pixels must be at least 20, and the
metal pattern must form a single connected region with a
continuous path back to the feed strip.
\par
Patterns are generated with equal quota per family to ensure
balanced topological coverage. All EM simulations were performed in CST\,Microwave\,Studio
using the time-domain solver over 22--30\,GHz, yielding the
complex $S_{11}(f)$ response sampled at 801 uniformly spaced
frequency points per design.
A Python script was used to interface with CST for geometry
generation, simulation, and result export.

\section{Machine Learning Modelling}
\label{sec:method}

\begin{figure}[!t]
  \centering
  \includegraphics[width=\columnwidth]{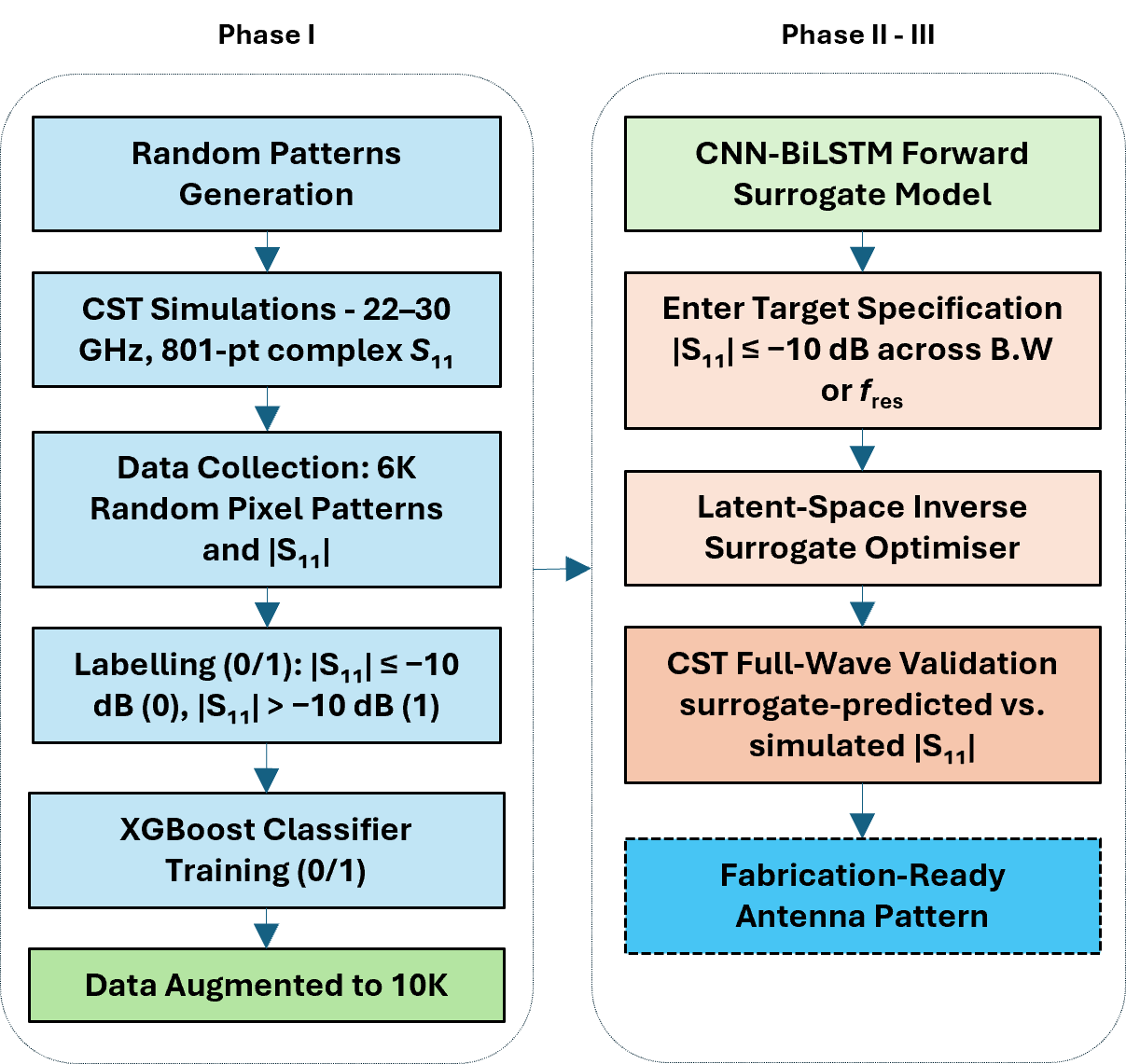}
  \caption{End-to-end system pipeline.
           An XGBoost classifier screens possible patterns to
           augment the CST dataset.
           A CNN-BiLSTM forward surrogate and a convolutional
           AE prior are trained on the augmented dataset.
           The inverse optimiser searches the 64-d AE latent
           space through the frozen surrogate to generate
           antenna patterns satisfying the target $S_{11}$
           specification.}
  \label{xgboost}
\end{figure}

As shown in Fig.~\ref{xgboost}, the ML modelling was
carried out in three phases: (i)~classifier-guided dataset
augmentation, (ii)~forward surrogate training, and
(iii)~latent-space inverse optimisation.

\subsection{Classifier-Guided Data Augmentation}
\label{sec:aug}

An initial set of approximately 6{,}000 pixel patterns was
generated using the structured-random strategy described in
Section~\ref{sec:structure} and simulated in CST.
Each pattern was labelled \emph{resonant} (positive class) if
its deepest point, $\min_f |S_{11}(f)|_{\mathrm{dB}}$, fell
below $-10$\,dB anywhere in 22--30\,GHz, and
\emph{non-resonant} (negative class) otherwise.
\par
Approximately 40\% of the initial patterns were resonant,
leaving the dataset heavily skewed toward non-resonant
designs. To improve the efficiency of further data collection, five
binary classifiers were evaluated to distinguish resonant from
non-resonant patterns prior to simulation.
A balanced subset of 4{,}178 samples (2{,}089 per class) was
used for training, with stratified 70/15/15
train/validation/test splits.
The models were evaluated using test accuracy, macro F1-score,
ROC\,AUC, and train-to-validation accuracy gap, as reported in
Table~\ref{tab:clf6k}.
XGBoost was selected for the augmentation step as it achieved
the highest test accuracy among models with a low
train-to-validation gap (77.4\%, $\Delta{=}4.4\%$),
indicating reliable generalisation to unseen pixel
patterns~\cite{cawley2010over}.

\begin{table}[!t]
  \caption{Classifier benchmark on the 6K balanced dataset.
           Best result per model across five random seeds.
           \textbf{Bold}: best per column.}
  \label{tab:clf6k}
  \centering
  \renewcommand{\arraystretch}{1.12}
  \small
  \setlength{\tabcolsep}{5pt}
  \begin{tabular}{lcccc}
    \toprule
    \textbf{Model} & \textbf{Acc.} & \textbf{F1} &
    \textbf{AUC} & \textbf{Train-Val Gap} \\
    \midrule
    \textbf{Random Forest} & \textbf{80.1\%} & \textbf{0.796} & \textbf{0.874} & 14.7\% \\
    XGBoost                & 77.4\%          & 0.768          & 0.844          & 4.4\%  \\
    LightGBM               & 77.2\%          & 0.772          & 0.839          & \textbf{3.4\%}  \\
    SVM                    & 75.9\%          & 0.761          & 0.827          & 4.1\%  \\
    Logistic Reg.          & 71.0\%          & 0.706          & 0.790          & 3.6\%  \\
    \bottomrule
  \end{tabular}
\end{table}
\vspace{-1mm}
\par Although Random Forest achieved higher accuracy, its large
train-to-validation gap of 14.7\% suggests overfitting that
would reduce its reliability as a screener on new unseen
patterns. \par 
The model used a maximum tree depth of 3, a learning rate of
0.1, and subsampling at the sample and feature level, with
the number of boosting rounds determined by early stopping on
the validation log-loss. The trained XGBoost classifier was then used to score a large
pool of newly generated random patterns and retain only those
predicted to be resonant for CST simulation.
An additional 4{,}000 patterns were selected and simulated in
this way, expanding the total dataset to approximately
10{,}000 samples.
The proportion of resonant designs in the combined dataset
increased from approximately 40\% to 52\%, confirming that
the classifier-guided selection meaningfully enhanced the
training dataset as compared to uniform random simulation.

\subsection{Forward Surrogate Model (CNN-BiLSTM)}
\label{sec:fwd}
\vspace{2 mm}
\subsubsection{Architecture}
\begin{figure}[!t]
  \centering
  \includegraphics[width=\columnwidth]{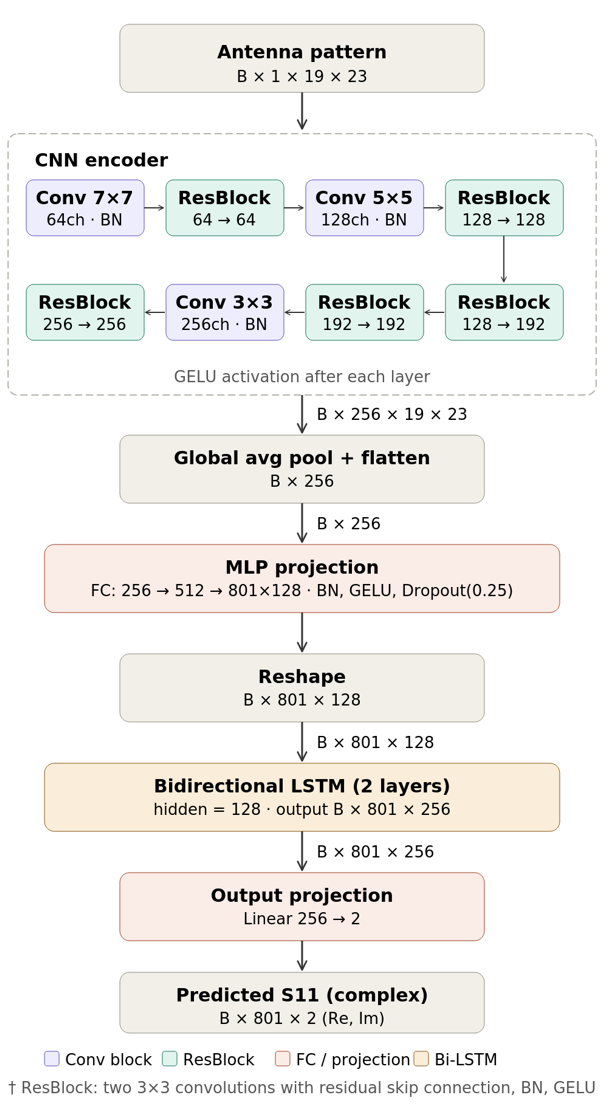}
  \caption{CNN-BiLSTM forward surrogate architecture.
           The spatial encoder extracts geometric features
           from the binary pixel pattern, the frequency
           projection head reshapes them into a per-frequency
           sequence, and the bidirectional LSTM models
           inter-frequency correlations to output the full
           complex $S_{11}$ spectrum.}
  \label{fig:pipeline}
\end{figure}

As shown in Fig.~\ref{fig:pipeline}, the forward surrogate maps a binary $19{\times}23$ pixel
pattern to the 801-point complex $S_{11}$ spectrum over
22--30\,GHz.
The network consists of: (i)~a \emph{convolutional spatial
encoder} comprising three Conv2D--BatchNorm--GELU blocks with
kernel sizes 7, 5, and 3, each followed by a residual block,
with channel depths progressing
$1{\to}64{\to}128{\to}192{\to}256$, followed by global
average pooling to produce a 256-dimensional feature vector;
(ii)~a \emph{frequency projection head} comprising two linear
layers (256\,${\to}$\,512\,${\to}$\,$801{\times}128$) with
BatchNorm, GELU, and Dropout ($p{=}0.25$) that reshapes the
output to an $801{\times}128$ sequence; and (iii)~a
\emph{2-layer bidirectional LSTM} that models inter-frequency
correlations and outputs a $801{\times}2$ real/imaginary
$S_{11}$ tensor.
Dropout ($p{=}0.25$) is applied within both the encoder and
the LSTM.

\vspace{2mm}
\subsubsection{Loss Function}
Training minimises a composite loss:
\begin{equation}
  \mathcal{L}
    = \lambda_1 \mathcal{L}_{\mathrm{Re/Im}}
    + \lambda_2 \mathcal{L}_{\mathrm{dB}}
    + \lambda_3 \mathcal{L}_{\mathrm{depth}}
    + \lambda_4 \mathcal{L}_{\mathrm{freq}}
    + \lambda_5 \mathcal{L}_{\mathrm{slope}},
  \label{eq:loss}
\end{equation}
where $\mathcal{L}_{\mathrm{Re/Im}}$ is the MSE on the
complex $S_{11}$; $\mathcal{L}_{\mathrm{dB}}$ is a Huber
loss ($\delta{=}3$\,dB) on the dB-magnitude;
$\mathcal{L}_{\mathrm{depth}}$ is an L1 loss on the predicted
resonance dip depth; $\mathcal{L}_{\mathrm{freq}}$ is an L1
loss on the predicted resonance frequency, computed via a
differentiable soft-argmin estimator; both
$\mathcal{L}_{\mathrm{depth}}$ and $\mathcal{L}_{\mathrm{freq}}$
are computed over resonant samples
(true $|S_{11}|_{\mathrm{dB}} < -10$\,dB); and
$\mathcal{L}_{\mathrm{slope}}$ is the MSE on the
first-order finite difference of the dB spectrum, which
penalises non-smooth predictions that would not occur in a real EM simulation.
The weights are set to
$(\lambda_1,\lambda_2,\lambda_3,\lambda_4,\lambda_5)
{=}(1.0,\,2.0,\,4.0,\,15.0,\,0.3)$.
\vspace{2mm}
\subsubsection{Training}
The full 10{,}000-sample dataset was split 70/15/15
(stratified, seed\,42) into approximately 7{,}000 training,
1{,}500 validation, and 1{,}500 test samples.
The model was trained for up to 200 epochs, with training
stopped early if no improvement was seen for 30 consecutive
epochs.
The AdamW optimiser was used with a peak learning rate of
$5{\times}10^{-5}$ and weight decay of $10^{-4}$.
The learning rate was increased linearly over the first 3
epochs before following a cosine decay schedule, which
prevents large weight updates at the start of training. \par
To reduce the effect of class imbalance, resonant samples
were drawn $4{\times}$ more frequently during training using
a weighted sampler.
Random horizontal and vertical flips of the pixel pattern
were also applied as data augmentation.

\subsection{Convolutional Autoencoder}
\label{sec:ae}

Direct optimisation over the full $19{\times}23$ binary pixelated antenna surface can produce disconnected or physically invalid patterns
that cannot radiate.
To address this, a convolutional autoencoder (AE) is trained
on all the pixelated antenna patterns to learn a compact
latent representation of feasible antenna geometries,
providing a structured search space for the inverse optimiser.
The encoder compresses a $1{\times}19{\times}23$ binary pattern
through three Conv2D blocks with channel depths
$1{\to}32{\to}64{\to}128$, where the second and third
convolutions use stride~2, reducing the spatial resolution
to a $128{\times}5{\times}6$ feature map that is then
projected to a 64-dimensional latent vector~$\mathbf{z}$.
The decoder reconstructs the pattern via bilinear upsampling
and convolution, outputting per-pixel logits.
\par The training loss combines binary cross-entropy with a
total-variation regulariser:
\begin{equation}
  \mathcal{L}_{\mathrm{AE}}
    = \mathcal{L}_{\mathrm{BCE}}
    + \lambda_{\mathrm{TV}}\,\mathcal{L}_{\mathrm{TV}},
  \quad \lambda_{\mathrm{TV}} = 10^{-3},
  \label{eq:ae}
\end{equation}
trained with AdamW ($\mathrm{lr}{=}10^{-3}$) for up to 150
epochs with early stopping (patience\,=\,15).

\subsection{Latent-Space Inverse Optimiser}
\label{sec:inv}

The inverse design problem is formulated as gradient descent
in the AE latent space.
Rather than optimising the 437 binary pixel values directly,
the search is carried out in the 64-dimensional latent space
of the trained AE, which constrains possible patterns to
physically valid antenna geometries and reduces the
dimensionality of the problem:
\begin{equation}
  \mathbf{z}^{\!*} = \arg\min_{\mathbf{z}}\;
    \mathcal{L}_{\mathrm{inv}}\!\bigl(
      f_{\mathrm{sur}}\!\bigl(
        \Pi(\sigma(g_{\mathrm{AE}}(\mathbf{z})))
      \bigr),\;
      \mathbf{y}_{\mathrm{tgt}}
    \bigr)
    + \mathcal{R}(\mathbf{z}),
  \label{eq:inv}
\end{equation}

where $g_{\mathrm{AE}}$ is the frozen AE decoder, $\sigma$ is
the sigmoid function, $\Pi$ is a connectivity projection
operator, $f_{\mathrm{sur}}$ is the frozen forward surrogate,
and $\mathcal{R}(\mathbf{z})$ combines $\ell_1$ sparsity,
total-variation, pixel-area, and $\ell_2$-latent
regularisers.
The target loss $\mathcal{L}_{\mathrm{inv}}$ penalises in-band
$|S_{11}|$ values above $-11$\,dB (1\,dB margin) and
suppresses spurious resonances outside a $\pm1$\,GHz guard
band around the target.

At every optimisation step, the continuous decoder output is
passed through a sigmoid and thresholded to produce a binary
pattern.
The binarised pattern then undergoes connected-component
labelling; only the largest metal region contiguous with the
feed is retained, ensuring every possible geometry is both
electrically active and fabrication-ready.
A straight-through estimator~\cite{bengio2013ste} passes
gradients through this discrete step so that the optimisation
can continue uninterrupted.
Optimisation runs for 1{,}000 steps using Adam with a cosine
annealing schedule ($\mathrm{lr}_0{=}0.05$).

To improve robustness, a $K{=}4$ dropout ensemble of surrogate
passes is used at each step, along with cyclic frequency
shifts ($\pm60$, $\pm120$\,bins) applied to the target band,
which reduces sensitivity to surrogate frequency bias.
A multi-start strategy is also employed in which a population
of latent vectors is independently initialised and optimised,
with half seeded from encoded resonant training patterns and
the remainder drawn from a standard normal distribution.
The possible antenna yielding the minimum in-band predicted
$|S_{11}|_{\mathrm{dB}}$ after convergence is selected as the
final design.

\section{Results}
\label{sec:results}

\subsection{Forward Surrogate Performance}

Table~\ref{tab:fwd} reports the CNN-BiLSTM surrogate
performance across the train, validation, and test splits of
the 10K dataset.
The broadband MAE and RMSE are consistent between validation
and test (1.55 and 2.74\,dB respectively), confirming that
the model generalises well to unseen patterns.
Resonant metrics are computed over the 52\% of samples
with a true $|S_{11}|$ minimum below $-10$\,dB.
The surrogate achieves a mean dip frequency MAE of
0.72\,GHz across the test set; the inverse optimiser's
multi-start selection strategy consistently favours
candidates where the surrogate prediction is most reliable,
yielding closer agreement in the validated designs shown below.

\begin{table}[!t]
  \caption{CNN-BiLSTM forward surrogate performance metrics}
  \label{tab:fwd}
  \centering
  \renewcommand{\arraystretch}{1.12}
  \small
  \begin{tabular}{lccc}
    \toprule
    \textbf{Metric} & \textbf{Train} & \textbf{Val} & \textbf{Test} \\
    \midrule
    MAE, $|S_{11}|_{\mathrm{dB}}$                        & 1.305 & 1.547 & 1.548 \\
    RMSE, $|S_{11}|_{\mathrm{dB}}$                       & 2.218 & 2.725 & 2.735 \\
    MAE, $f_{\mathrm{res}}$ depth (dB)\textsuperscript{*}         & 3.213 & 5.844 & 5.742 \\
    MAE, $f_{\mathrm{res}}$ centre frequency (GHz)\textsuperscript{*} & 0.576 & 0.681 & 0.724 \\
    \bottomrule
  \end{tabular}
\end{table}

\subsection{Inverse Design Results}

To validate the full pipeline, four antenna patterns were
generated by the inverse optimiser, each targeting a different
frequency band within 22--30\,GHz.
Each pattern was exported to CST and simulated using the
full-wave time-domain solver to verify the surrogate
prediction.
A design is considered successful if the CST-simulated
$|S_{11}|$ reaches $-10$\,dB or below within the target band.

\begin{figure}[!t]
  \centering
  \includegraphics[width=\columnwidth]{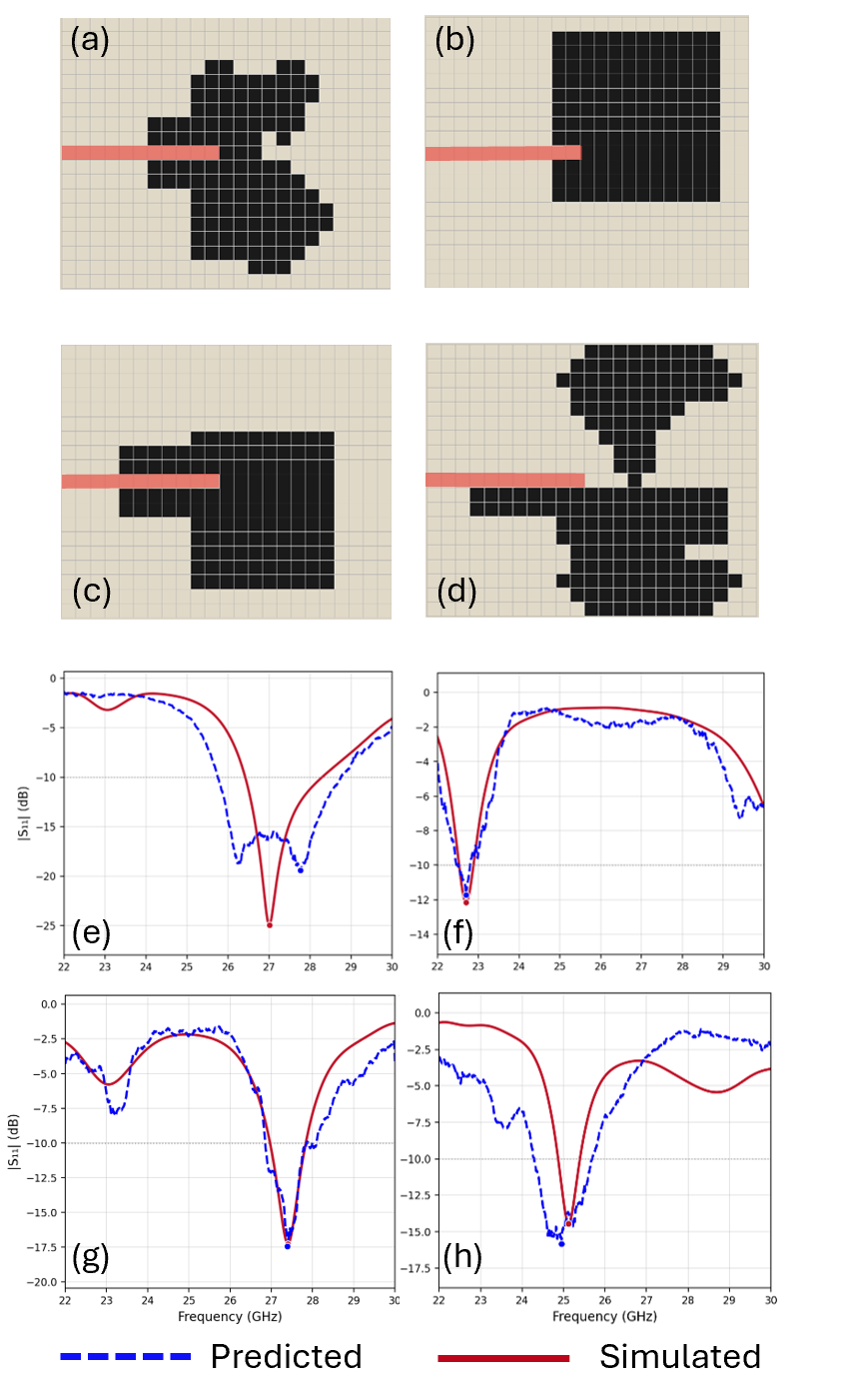}
  \caption{Inverse-designed pixelated patch antennas and
           their $|S_{11}|$ responses.
           (a)--(d) optimised pixel patterns for target bands
           26.5--27.5, 22--23, 27--28, and 25\,GHz
           respectively.
           (e)--(h) corresponding surrogate-predicted
           (dashed) vs.\ CST-simulated (solid)
           $|S_{11}|_{\mathrm{dB}}$ responses.}
  \label{fig:results}
\end{figure}

For the first case, a target $f_{\mathrm{res}}$ range of
26.5--27.5\,GHz was specified.
The inverse-optimised pixel pattern is shown in
Fig.~\ref{fig:results}a, while the surrogate-predicted and
CST-simulated $|S_{11}|_{\mathrm{dB}}$ responses are compared
in Fig.~\ref{fig:results}e.
The predicted centre $f_{\mathrm{res}}$ was 27.76\,GHz,
while the CST-simulated centre $f_{\mathrm{res}}$ was
27\,GHz, giving a frequency error of 0.76\,GHz and
$\Delta|S_{11}| = 5.5$\,dB.

For the second case, a target $f_{\mathrm{res}}$ range of
22--23\,GHz was specified.
The pixel pattern and $|S_{11}|$ comparison are shown in
Figs.~\ref{fig:results}b and~\ref{fig:results}f.
Both the predicted and CST-simulated centre $f_{\mathrm{res}}$
were 22.7\,GHz, with $\Delta|S_{11}| = 0.44$\,dB, showing
close agreement between the surrogate and simulation.

For the third case, a target $f_{\mathrm{res}}$ range of
27--28\,GHz was specified.
The pixel pattern and $|S_{11}|$ comparison are shown in
Figs.~\ref{fig:results}c and~\ref{fig:results}g.
The predicted centre $f_{\mathrm{res}}$ was 27.38\,GHz and
the CST-simulated centre $f_{\mathrm{res}}$ was 27.4\,GHz,
giving $\Delta|S_{11}| = 0.18$\,dB.

For the fourth case, a target $f_{\mathrm{res}}$ of 25\,GHz
was specified.
The pixel pattern and $|S_{11}|$ comparison are shown in
Figs.~\ref{fig:results}d and~\ref{fig:results}h.
The predicted centre $f_{\mathrm{res}}$ was 24.95\,GHz and
the CST-simulated centre $f_{\mathrm{res}}$ was 25.12\,GHz,
giving $\Delta|S_{11}| = 1.4$\,dB. \par
Across the four designs, three show frequency errors below
0.2\,GHz and depth errors below 1.5\,dB, confirming that the
surrogate provides a reliable gradient signal for inverse
optimisation.
The first design shows a larger depth discrepancy of 5.5\,dB,
which is consistent with the test-set dip depth MAE of
5.74\,dB reported in Table~\ref{tab:fwd}.
All four CST-simulated patterns achieved $|S_{11}| \leq
-10$\,dB within the specified target band, satisfying the
design objective in every case.

The results show that the pipeline consistently produces connected, feed-attached pixel patterns that resonate
within the specified target band across the 22--30\,GHz range.
The bidirectional LSTM and the composite
$\mathcal{L}_{\mathrm{depth}}$/$\mathcal{L}_{\mathrm{freq}}$ loss terms were found to be key contributors to surrogate
accuracy, while connectivity projection at every optimisation
step ensured that all generated patterns maintain a continuous
electrical path to the feed.

\section{Conclusion and Future Work}
\label{sec:conc}

In this paper, a novel surrogate-assisted inverse design
process for pixelated mmWave patch antennas was proposed.
An XGBoost binary classifier was used as a pre-simulation
filter for dataset augmentation, raising the proportion of
resonant designs from approximately 40\% to 52\% across a
combined dataset of ${\sim}10{,}000$ simulations.
A CNN-BiLSTM forward surrogate trained on this dataset
predicts the full complex $S_{11}$ spectrum at 801 frequency
points over 22--30\,GHz, guided by a physics-aware composite
loss.
A 64-dimensional convolutional AE prior constrains the inverse
search to the manifold of feasible antenna geometries;
gradient descent through the frozen surrogate with
connectivity projection yields fabrication-ready pixel
patterns satisfying a target $|S_{11}| \leq -10$\,dB
specification.
Four inverse-designed geometries were presented and validated
against full-wave simulation. It was shown, that the proposed approach enables to automatically design and configure effective custom antenna structures.

Future work will proceed along three directions.
First, the inverse-designed prototypes will be fabricated and
measured.
This will quantify the simulation-to-hardware gap and
establish the practical accuracy limits of the surrogate
model under real fabrication tolerances. Second, the framework will be extended to incorporate radiation pattern, gain, and
efficiency as additional design targets alongside $S_{11}$.
Third, the pipeline will be extended to two-port bandpass
filter synthesis using the same pixelated topology, requiring
the surrogate to predict the full scattering
matrix and the inverse loss to enforce prescribed passband
insertion loss and stopband rejection simultaneously.

\section*{Acknowledgment}
This publication has emanated from research conducted with the financial support of Research Ireland under Grant number \text{13/RC/2077\_P2}. For the purpose of Open Access, the author has applied a CC BY public copyright licence to any Author Accepted Manuscript version arising from this submission.

\bibliographystyle{IEEEtran}
\bibliography{refs.bib}

\vfill
\end{document}